\documentclass[%
 reprint,
longbibliography,
nofootinbib,
 amsmath,amssymb,
 aps,
]{revtex4-2}

\usepackage{times}
\usepackage{newtxtext,newtxmath}

\usepackage{adjustbox}
\usepackage{xcolor}
\usepackage{mathtools}
\usepackage{braket}
\usepackage{soul}
\usepackage{amsmath}
\usepackage{multirow}
\usepackage{relsize}
\usepackage{tabularx}
\usepackage{array}
\usepackage{soul}
\usepackage{graphicx}
\usepackage{dcolumn}
\usepackage{bm}
\usepackage{hyperref}
\usepackage{ upgreek }

\begin{document}

\preprint{APS/123-QED}

\title{Non-Markovian dynamics in nonstationary Gaussian baths: a hierarchy of pure states approach}

\author{Vladislav Sukharnikov}%
\email{vladislav.sukharnikov@desy.de}
\affiliation{
Universität Hamburg, Luruper Chaussee 149, 22761 Hamburg, Germany
}%

\affiliation{
The Hamburg Centre for Ultrafast Imaging, Luruper Chaussee 149, 22761 Hamburg, Germany
}

\author{Stasis Chuchurka}%

\affiliation{
Universität Hamburg, Luruper Chaussee 149, 22761 Hamburg, Germany
}%

\author{Frank Schlawin}%
\email{frank.schlawin@mpsd.mpg.de}
\affiliation{Max Planck Institute for the Structure and Dynamics of Matter, Luruper Chaussee 149, 22761 Hamburg, Germany
}%
\affiliation{
Universität Hamburg, Luruper Chaussee 149, 22761 Hamburg, Germany
}%
\affiliation{
The Hamburg Centre for Ultrafast Imaging, Luruper Chaussee 149, 22761 Hamburg, Germany
}

\begin{abstract}
    Building on the standard hierarchy of pure states (HOPS) approach, we construct a generalized formulation suitable for open quantum systems interacting with nonstationary Gaussian baths, potentially extending its applicability to nonequilibrium baths. This is achieved by extending the conventional exponential decomposition of a bath correlation function (BCF) for nonstationary cases. Using our formulation of HOPS, we derive the corresponding hierarchy of master equations and, when each term in the BCF expansion can be associated with an independent physical bath, we show how the formalism connects to the well-known pseudomode representation. We demonstrate the method's performance on two examples of nonstationary squeezed reservoirs generated via uniform squeezing and degenerate parametric amplification in a one-sided cavity. Benchmarking against the hierarchy of master equations shows that HOPS is more efficient under hierarchy truncation. The pseudomode representation is shown to be more efficient in the strongly non-Markovian regime. Our results highlight HOPS as a versatile and powerful tool for simulating open quantum systems in nonstationary baths, with potential applications ranging from squeezed light-matter interactions to driven quantum materials and dissipative phase transitions.
\end{abstract}

\maketitle

\section{\label{sec: introduction} Introduction}

Every physical system is coupled to its surroundings, which can lead to the loss of quantum coherence and unavoidable increase of entropy. Thus, it is often necessary to treat these systems as open quantum systems. The surrounding is usually considered to be in a thermal state---a common assumption for the overwhelming majority of cases. However, the development of strong lasers in the mid-IR and terahertz regime has led to the emergence of a new field of research in quantum materials, where strong excitation of high-energy degrees of freedom is used to manipulate and control the low-energy physics of these materials~\cite{SentefRMP}. 
For instance, in the field of nonlinear phononics, a specific optically active phonon mode is driven to large amplitudes~\cite{Foerst2011, Mankowsky_2016, Subedi2021}. Through nonlinear coupling to other vibrations, this excitation may be downconverted to transiently change the lattice structure~\cite{Disa2021} and affect the energetic balance between competing low-energy phases. 
These and similar approaches are successfully used to stabilize coherent phases such as superconductivity~\cite{Mankowsky2014, Budden2021}, magnetic ordering~\cite{Disa2023}, or ferroelectricity~\cite{Nova2019} above their equilibrium critical temperature, or even to induce transient phases that are not found in the material at equilibrium~\cite{Buzzi2020, Tindall2020, Tindall2021,Tindall2021liebstheorem}. 
Although in the theory accompanying these seminal experiments the phonon modes are typically treated classically, their action on the low-energy degrees of freedom should be more appropriately modeled as a bath that is driven far from equilibrium.

A similar situation may be found in photoinduced chemical reactions, where, e.g., the excitation of phonon modes is used to control the charge transfer reaction that it mediates~\cite{Delor2015, Valianti2019}. 
In dissipative phase transitions, this is even more explicit~\cite{MULLER2012, Sieberer_2016, Debecker2024}, since its goal is to manipulate the coupling to a bath so that the bath drives the system into a dark state with desirable properties, such as long-range coherence. 

These developments motivate us to investigate the quantum dynamics of open systems interacting with baths out of equilibrium. A fully general description remains challenging. We thus hypothesize that, in many situations, the system's dynamics can be approximated by linear coupling to a nonstationary Gaussian bath. When only the system is of interest, the bath need not be tracked explicitly: its degrees of freedom can be integrated out, yielding time-nonlocal (Nakajima-Zwanzig) or time-convolutionless master equations \cite{Breuer2007, Nakajima1958, Zwanzig1960, deVega2017}, which, however, are difficult to use in practice. More tractable equations can be obtained by introducing approximations, such as the Markovian approximation or the assumption of weak system–bath coupling.

Nonetheless, these approximations are not generally applicable, and several methods have been developed to go beyond their limits. For instance, the reduced dynamics can be formulated in terms of the bath-correlation function (BCF). For a linearly coupled Gaussian bath, this permits genuinely nonperturbative non-Markovian treatments, since the reduced dynamics is fully characterized by the mean bath field and its BCF \cite{Tamascelli2018}. For example, the non-Markovian quantum-state diffusion approach employs the BCF \cite{Disi1997, Disi1998}, though its practical use is limited. A widely used framework is the hierarchical equations of motion (HEOM), which expands the BCF as a sum of exponentials, enabling a representation of non-Markovian dynamics via auxiliary density operators \cite{Tanimura1989, Tanimura2006}. This approach is in a similar spirit to the pseudomode approach \cite{Garraway1997, Stenius1996, Imamoglu1994, Pleasance2020}, which approximates the coupling to a bath with a continuous spectrum by interactions with a finite set of effective broadband modes subject to Markovian decoherence. Related approaches reconstruct the BCF with networks of coupled auxiliary bosons \cite{Mascherpa2020}, map the bath to chains of effective modes \cite{Hughes2009, Hughes2009_II}, extend the pseudomode approach via quasi-Lindblad generators \cite{Park2024}, and generalize HEOM and pseudomode frameworks to tackle computationally demanding regime of ultra-strong coupling \cite{Lambert2019}.

Most such methods evolve density operators. However, it is often more efficient to sample stochastic pure-state trajectories and reconstruct the density matrix via ensemble averaging \cite{Disi1998, Strunz1999}. A notable method in this direction is the hierarchy of pure states (HOPS) approach \cite{Suess2014, Hartmann2017, Hartmann2021, Gera2023}, which assumes exponential decomposition of the BCF and unravels non-Markovian dynamics into a set of stochastic differential equations for auxiliary states. This method reduces the dimensionality of the state space from Liouville space to Hilbert space and, in some cases, has been shown to converge faster with respect to the number of hierarchy levels than approaches based on hierarchies of density operators \cite{Zhang2016}.

At thermal equilibrium, the BCF is stationary---i.e., invariant under time translations \cite{Breuer2007}. Out of equilibrium or under external driving, the BCF can become explicitly time-dependent and thus loses time-translation invariance. The original HOPS formulation assumes a stationary bath, but there are extensions to nonstationary squeezed reservoirs \cite{Link2022} that yield time-local hierarchical equations for auxiliary density operators derived from HOPS. However, a systematic and efficient stochastic pure-state treatment for broader classes of nonstationary baths is still lacking. In this article, we derive and benchmark a more general HOPS formulation for explicitly nonstationary baths. We assume a specific ansatz for the BCF and build the HOPS formulation in a pseudo-Fock space \cite{Gao2022}. In this representation, the open quantum system interacts with a finite set of effective modes, and nonstationarity renders the effective Hamiltonian explicitly time-dependent.

This article is organized as follows. In Sec. \ref{sec: OQS}, we develop a nonstationary extension of the hierarchy of pure states (HOPS) that treats explicitly time-dependent Gaussian baths while retaining the favorable scaling of standard HOPS. From this framework, we additionally derive two complementary formulations: a hierarchy of master equations, and, when each effective mode of the bath can be assigned to an independent bath, a pseudomode representation. We benchmark these methods in Sec. \ref{sec: numerical} on two models of squeezed reservoirs, specifically the uniformly squeezed vacuum of Ref. \cite{Link2022} and the output of the degenerate parametric amplifier \cite{Gardiner1984, Collett1984}. These two models provide a minimal, analytically controlled route to explicit time dependence in Gaussian baths and span non-Markovian to near-Markovian regimes. Finally, in Sec. \ref{sec: discussion}, we show how a finite-temperature displaced–squeezed thermal bath can be embedded in our formalism.

\section{Theory}\label{sec: OQS} 

We consider a quantum system coupled to a bath composed of a continuum of bosonic modes. In the interaction picture with respect to the bath, the Hamiltonian is
\begin{equation}\label{eq: Hamiltonian general}
    \hat{H}(t) = \hat{H}_S(t) + \hbar \hat{L}  \hat{B}(t),
\end{equation}
where $\hat{H}_S(t)$ is the Hamiltonian of the isolated system, which may be time-dependent. The second term describes system-bath coupling: $\hat{L}$ and $\hat{B}(t)$ are Hermitian operators of the system and the bath, respectively. 

Simulating the full system is generally intractable due to the infinite number of bath degrees of freedom. Instead, we focus on the system dynamics by tracing out the bath, assuming: (i) the system and bath are initially uncorrelated; (ii) the bath starts in a Gaussian state $\hat{\rho}_B$; and (iii) $\hat{B}(t)$ is linear in the bath's bosonic operators. Under these conditions, the bath's influence on the reduced system dynamics is fully determined by the mean bath field $\mathrm{Tr}[\hat{B}(t) \hat{\rho}_B]$ and the bath correlation function (BCF)  \cite{Tamascelli2018, Mascherpa2020}
\begin{equation}\label{eq: BCF introduction}
    \alpha(t,s) = \mathrm{Tr} [ \hat{B}(t) \hat{B}(s)  \hat{\rho}_B ],
\end{equation}
which satisfies $\alpha^*(t,s) = \alpha(s,t)$. Any pair of a linear $\hat{B}(t)$ and a Gaussian initial state that reproduces these two quantities results in the same reduced system dynamics. We assume that $\mathrm{Tr} [ \hat{B}(t) \hat{\rho}_B ] = 0$ without loss of generality, since the mean field may always be absorbed into the Hamiltonian.

We are interested in nonstationary baths, where the correlation function depends on both times explicitly, namely $\alpha(t,s) \not = \alpha(t-s)$. In what follows, we show how existing hierarchical methods can be adapted to such cases with only minor modifications.

\subsection{\label{sec: HOPS} Hierarchy of pure states (HOPS)}

When the bath is stationary and characterized by a time-translation invariant BCF, $\alpha(t,s) = \alpha(t - s)$, the reduced system dynamics can be efficiently treated using the hierarchy of pure states (HOPS) \cite{Suess2014}. In this method, the continuous bath spectrum is approximated by a finite set of effective broadband modes. The reduced system dynamics is described by averaging over stochastic states, rather than propagating density operators of larger dimension.

The original HOPS formalism applies when the BCF can be approximated by a sum of exponentials:
\begin{equation}\label{eq: stationary BCF}
    \alpha(t-s) = \sum_{j=1}^{N} \gamma_j \alpha_j(t-s) e^{-i\omega_j(t-s)},
\end{equation}
where each $j$ labels an effective mode with frequency $\omega_j$, weighted by a complex coefficient $\gamma_j$. The kernels $\alpha_j(\tau)$ have an exponential form:
\begin{equation}\label{eq: exponential kernel}
    \alpha_j(\tau) = \dfrac{\Gamma_j}{2} e^{-\Gamma_j|\tau|},
\end{equation}
where the positive parameter $\Gamma_j$ denotes the inverse memory time, or the spectral half-width of the $j$th mode. In the Markovian limit, $\Gamma_j \rightarrow \infty$, the kernel approaches a delta function, indicating negligible memory effects.

Extension of HOPS to nonstationary BCFs has recently been explored in Ref. \cite{Link2022}, which considered a specific BCF for a squeezed reservoir. In this article, we consider a more general class of nonstationary BCFs, $\alpha(t,s) \neq \alpha(t-s)$, of the form
\begin{equation}\label{eq: BCF ansatz}
    \alpha(t,s) = \sum_{j=1}^{N} \alpha_j(t-s) f_j(t) g_j^*(s).
\end{equation}
The functions $f_j(t)$ and $g_j(s)$ introduce explicit time dependence. They cannot be chosen independently; there is an implicit relation between them due to the Hermiticity and positive semidefiniteness of the BCF.

In Appendix~\ref{sec: Appendix A}, we provide our derivation of the hierarchy of pure states, adapted for the BCF in Eq.~\eqref{eq: BCF ansatz}. We use the pseudo-Fock basis \cite{Gao2022} to obtain a compact form of the equations of motion, and present only the final equations below.

The system's reduced density matrix $\hat{\rho}_S(t)$ is obtained as a statistical average over stochastic pure states $|\psi(t)\rangle$:
\begin{equation}\label{eq: reduced density matrix sampling}
    \hat{\rho}_S(t)
    =
    \mathbb{E}\big[|\psi(t)\rangle\langle \psi(t)|\big].
\end{equation}
The statistical properties of the initial state $|\psi(0)\rangle$ are chosen such that the average reproduces $\hat{\rho}_S{(0)}$. System observables are computed analogously:
\begin{equation}\label{eq: expectation values}
   \mathrm{Tr}\big[ \hat{O} \hat{\rho}_S(t) \big]
   =
   \mathbb{E}\big[\langle \psi(t) | \hat{O} |\psi(t)\rangle \big],
\end{equation}
where $\hat{O}$ denotes an operator acting on the system. Reformulating the problem in terms of stochastic pure states reduces the dimensionality of the state space. This partially mitigates the bottleneck associated with the unfavorable scaling of Liouville space by trading it for statistical sampling, which can be easily parallelized. The main practical limitation instead lies in the convergence of the statistical averages, which is addressed later.

Simulating the dynamics of $|\psi(t)\rangle$ requires a set of auxiliary states $|\psi^{(\mathbf{n})}(t)\rangle$, commonly called a hierarchy \cite{Suess2014}. Each auxiliary state is indexed by a vector of $N$ non-negative integers, $\mathbf{n} = (n_1, \dots, n_j, \ldots, n_N)$ \cite{Hartmann2017}, where each index $j$ corresponds to a mode in the BCF ansatz \eqref{eq: BCF ansatz}. 

The elements of $\mathbf{n}$ can be interpreted as occupation numbers of effective modes, forming a pseudo-Fock space~\cite{Gao2022}. In this interpretation, the auxiliary states are projections of an extended state vector $|\Psi(t)\rangle$ onto a pseudo-Fock basis: $|\psi^{(\mathbf{n})}(t)\rangle = \langle \mathbf{n} | \Psi(t)\rangle$~\cite{Gao2022}. The physical state $|\psi(t)\rangle$, used to compute observables via Eq.~\eqref{eq: expectation values}, corresponds to the projection onto the vacuum state: 
\begin{equation}
    |\psi(t)\rangle = |\psi^{(\mathbf{0})}(t)\rangle = \langle \mathbf{0} | \Psi(t)\rangle.
\end{equation}
Truncating the hierarchy is equivalent to truncating the pseudo-Fock basis. The physical state is normalized on average, whereas the norms of the auxiliary states encode normal-ordered expectation values of the bath operator~\cite{Boettcher2024}.

The key theoretical result of our work is a stochastic Schrödinger equation for the extended state vector
\begin{equation}\label{eq: SSE (I)}
   \dfrac{\partial |\Psi(t)\rangle}{\partial t} 
   =\Big[
   -\sum_{j=1}^{N} \Gamma_j \hat{c}_j^\dag \hat{c}_j
   -\dfrac{i}{\hbar} \hat{H}_{\text{eff}}(t) 
   -iZ^*(t) \hat{L} 
   \Big] |\Psi(t)\rangle.
\end{equation}
Here, $\hat{c}_j$ and $\hat{c}^\dag_j$ are bosonic operators defined in the pseudo-Fock space, satisfying $[\hat{c}_i, \hat{c}_j^\dag] = \delta_{ij}$. Each bosonic mode corresponds to an effective mode in the decomposition of the BCF in Eq. \eqref{eq: BCF ansatz}. The first term describes damping of these modes, with decay rates $\Gamma_j$ specified in Eq. \eqref{eq: exponential kernel}.

The second term in Eq.~\eqref{eq: SSE (I)} contains the effective Hamiltonian $\hat{H}_{\text{eff}}(t)$, which includes the isolated system Hamiltonian $\hat{H}_S(t)$ and interaction with the effective modes:
\begin{equation}\label{eq: effective Hamiltonian}
     \hat{H}_{\text{eff}}(t)
     =
     \hat{H}_S(t)
     + \hbar \sum_{j=1}^{N} \sqrt{\dfrac{\Gamma_j}{2}} \big\{ f_j(t) \hat{c}_j + g^*_j(t) \hat{c}_j^\dag \big\} \hat{L}.
\end{equation}
The functions $f_j(t)$ and $g_j(t)$ originate from the BCF~\eqref{eq: BCF ansatz} and introduce time dependence into the coupling terms. In general, $\hat{H}_{\text{eff}}(t)$ is non-Hermitian, except when $f_j(t) = g_j(t)$. 

The last term in Eq. \eqref{eq: SSE (I)} involves a complex-valued stochastic process $Z^*(t)$ with zero mean and correlations determined by the BCF:
\begin{equation}\label{eq: noise autocorrelation}
    \mathbb{E}[ Z(t) Z^*(s)] = \alpha(t,s),
    \quad\quad
    \mathbb{E}[Z(t) Z(s)] = 0.
\end{equation}
Appendix \ref{Sampling Z} outlines the sampling procedure for $Z(t)$ via diagonalization of the BCF matrix. Notably, in the special case where $f_i(t)=g_i(t)$, the noise can be represented as a sum of independent Ornstein-Uhlenbeck processes.

The auxiliary states are initially unoccupied; hence, $|\Psi(0)\rangle$ is the tensor product of the system's initial state $|\psi(0)\rangle$ and the vacuum state in the pseudo-Fock space:
\begin{equation}\label{eq: initial condition pure states}
    |\Psi(0)\rangle = |\psi(0)\rangle \otimes |\mathbf{0}\rangle.
\end{equation}

Finally, we note that the standard HOPS formulation in pseudo-Fock space \cite{Gao2022} is recovered when
\begin{equation*}
    f_j(t) = \sqrt{\gamma_j} e^{-i\omega_j t},
    \quad\quad
    g_j(t) = \sqrt{\gamma_j^*} e^{-i\omega_j t}.
\end{equation*}
This corresponds to the stationary BCF described in Eq.~\eqref{eq: stationary BCF}.

\subsection{Nonlinear stochastic Schrödinger equation}\label{sec: nonlinear HOPS}

Monte Carlo evaluation of observables using Eq. \eqref{eq: expectation values} can converge slowly when using the solutions of the linear equation \eqref{eq: SSE (I)}, especially in the strong-coupling regime \cite{Hartmann2017}. This can be improved with a Girsanov transformation \cite{Suess2014, Hartmann2017, Disi1998}. First, the stochastic state's fluctuating norm is absorbed into the sampling measure, so observables are computed from normalized quantum trajectories. Second, the sampling is guided toward physically relevant regions where the state is localized \cite{Disi1998}. 

The reduced density matrix then takes the form
\begin{equation*}
    \hat{\rho}_S(t)
    =
    \mathbb{E}\bigg[\frac{|\tilde{\psi}(t)\rangle \langle \tilde{\psi}(t)|}{\langle \tilde{\psi}(t) | \tilde{\psi}(t) \rangle}\bigg],
\end{equation*}
with an analogous modification for the expressions of observables. The physical state vector is obtained via projection as $ |\tilde{\psi}(t)\rangle = \langle \mathbf{0} |\tilde{\Psi}(t)\rangle$. The state vector $|\tilde{\Psi}(t)\rangle$ evolves under a nonlinear stochastic equation
\begin{multline}\label{eq: nonlinear SSE (I)}
    \dfrac{d|\tilde{\Psi}(t)\rangle }{dt}
    = \Big[
    -\sum_{j=1}^{N} \Gamma_j \hat{c}_j^\dag \hat{c}_j
    -\dfrac{i}{\hbar} \hat{H}_{\text{eff}}(t)
    \\ -i\tilde{Z}^*(t) \hat{L}
    +iL(t)\sum_{j=1}^{N}\sqrt{\dfrac{\Gamma_j}{2}} f_j(t) \hat{c}_j
    \Big] |\tilde{\Psi}(t)\rangle.
\end{multline}
The initial condition $|\tilde{\Psi}(0)\rangle$ is unchanged by the Girsanov transformation and coincides with Eq.~\eqref{eq: initial condition pure states}. 

The nonlinearity in Eq. \eqref{eq: nonlinear SSE (I)} originates partly from the last term, which involves the normalized expectation value of $\hat{L}$ for a given realization of $|\tilde{\psi}(t)\rangle$:
\begin{equation}
    L(t) = \dfrac{\langle \tilde{\psi}(t) | \hat{L} | \tilde{\psi}(t)\rangle}{\langle \tilde{\psi}(t) |\tilde{\psi}(t)\rangle}.
\end{equation}
Additional nonlinearity arises from the modified noise term. Rather than the original $Z(t)$, the equation now involves
\begin{equation}\label{eq: tilde Z}
    \tilde{Z}(t) = Z(t) - i\! \int_0^t\!\! ds \, \alpha(t,s) L(s),
\end{equation}
which includes an additional term that can be viewed as an analogue of the conditional polarization field generated by the system for a given stochastic realization. Consequently, the noise ${Z}(t)$ appearing in the first term---whose correlation properties remain unchanged---represents fluctuations about this conditional field. The integral can be efficiently computed by solving a system of differential equations. In the special case where $f_j(t) = g_j(t)$, the entire $\tilde{Z}(t)$ can be obtained by solving a set of stochastic differential equations.

In the numerical simulations that follow, observables are computed from solutions of the nonlinear equation \eqref{eq: nonlinear SSE (I)}. The linear equation \eqref{eq: SSE (I)}, however, remains valuable, as it provides a route to an equivalent deterministic method---the hierarchy of master equations---which is also used in the numerical demonstrations.

\subsection{Hierarchy of master equations (HME)}
\label{sec: HEOM}

A deterministic formulation, complementary to the stochastic pure-state approach, can be obtained by averaging the projectors $|\Psi(t)\rangle\langle \Psi(t)|$, which yields the density operator
\begin{equation}\label{eq: extended density operator}
    \hat{\rho}(t) = \mathbb{E} \big[|\Psi(t)\rangle \langle \Psi(t)|\big].
\end{equation}
The reduced density matrix of the system is found by projection onto the vacuum state of the effective modes, $\hat{\rho}_S(t) = \langle \mathbf{0} | \hat{\rho}(t) | \mathbf{0} \rangle$. Therefore, the system's dynamics can be obtained by propagating $\hat{\rho}(t)$, starting from the initial state
\begin{equation}
\label{eq: HEOM initial condition}
    \hat{\rho}(0) = \hat{\rho}_S(0) \otimes |\mathbf{0}\rangle\langle\mathbf{0}|.
\end{equation}
This deterministic treatment avoids stochastic sampling but increases the size of the state space. The density operator $\hat{\rho}(t)$ evolves according to the following master equation:\footnote{This derivation uses the relation $ \mathbb{E} \big[ Z(t) |\Psi(t)\rangle \langle \Psi(t) | \big] = \displaystyle\sum_{j=1}^{N} \sqrt{\frac{\Gamma_j}{2}} f_j(t) \hat{c}_j \hat{\rho}(t)$ and its Hermitian conjugate.}
\begin{multline}\label{eq: HEOM}
    \dfrac{d\hat{\rho}(t)}{dt}
    =
    - \sum_{j=1}^{N} \Gamma_j \big\{ \hat{c}_j^\dag \hat{c}_j \hat{\rho}(t) + \hat{\rho}(t) \hat{c}_j^\dag \hat{c}_j \big\}
    \\
    -\dfrac{i}{\hbar} \hat{H}_\text{eff}(t) \hat{\rho}(t)
    +
    \dfrac{i}{\hbar} \hat{\rho}(t) \hat{H}^\dag_\text{eff}(t)
    \\
    -i\sum_{j=1}^{N} \sqrt{\dfrac{\Gamma_j}{2}} \big\{ f_j^*(t) \hat{L} \hat{\rho}(t) \hat{c}_j^\dag - f_j(t) \hat{c}_j \hat{\rho}(t) \hat{L} \big\},
\end{multline}
where $\hat{H}_\text{eff}(t)$ is given in Eq. \eqref{eq: effective Hamiltonian}. The equations for auxiliary density operators, $\langle \mathbf{n} | \hat{\rho}(t) |\mathbf{m}\rangle$, are known as a hierarchy of master equations (HME) \cite{Suess2015}. As shown in Ref. \cite{Suess2015}, this hierarchy coincides in certain cases with the hierarchical equations of motion originally introduced in Ref. \cite{Tanimura1989}.

As a side remark, the master equation \eqref{eq: HEOM} can, in principle, be unraveled using stochastic states driven by white noise, provided one enlarges the state space by allowing the stochastic bra and ket to differ and obey different equations. Related approaches have been considered in Refs. \cite{Menczel2024, Koch2008, Breuer1999}. However, for the problems considered in Sec. \ref{sec: numerical}, we found that the resulting equations converge poorly. In addition, when we normalized individual trajectories, some realizations became unstable and exhibited unbounded growth, and we were unable to stabilize the dynamics. For these reasons, we do not pursue this approach further here and provide no equations or figures. This behavior is reminiscent of the runaway trajectories encountered within the positive-$P$ representation \cite{Gilchrist1997, Deuar2021, compact_systems, XLO, hermitian}. This comparison highlights an advantage of the colored-noise unraveling provided by HOPS in Eq.~ \eqref{eq: nonlinear SSE (I)}: it evolves a single stochastic vector (the bra and ket are conjugate) and remains stable.

\subsection{Pseudomode master equation (PME)}\label{sec: pseudomode master equation}

In the special case where $f_j(t) = g_j(t)$, the master equation \eqref{eq: HEOM} can be rewritten in a more conventional form, consisting of a unitary evolution term together with a Lindblad dissipator acting on the effective modes. This reformulation is connected to the pseudomode representation \cite{Garraway1997}, where the open-system dynamics is described in terms of interactions with a finite set of bosonic modes subject to Markovian decoherence. Such a representation is particularly useful when the dissipation associated with $\Gamma_j$ is weak compared to the unitary dynamics, as illustrated in the numerical demonstrations. When the BCF is stationary as in Eq. \eqref{eq: stationary BCF}, the pseudomode representation exists when all $\gamma_j$ are real and positive.

The pseudomode formulation can be directly derived from the hierarchy of master equations by applying the following transformation to $\hat{\rho}(t)$ from Eq. \eqref{eq: extended density operator}:
\begin{equation}\label{eq: rotation transformation}
    \hat{\rho}'(t)
    =
    \mathrm{exp}\Big[ -\sum_{j=1}^{N} \hat{c}_j^{L} \hat{c}_j^{\dag R} \Big] \hat{\rho}(t).
\end{equation}
Here, the superscripts $L$ and $R$ denote left and right operator actions, respectively: $\hat{c}_j^L \hat{\rho} = \hat{c}_j \hat{\rho}$ and $\hat{c}_j^{\dag R} \hat{\rho} = \hat{\rho} \hat{c}_j^\dag$.

The system’s reduced density matrix is obtained by tracing $\hat{\rho}'(t)$ over the full pseudo-Fock basis:
\begin{equation}\label{eq: physical density matrix via pseudomode}
    \hat{\rho}_S(t) = \sum_{\mathbf{n}} \langle \mathbf{n}| \hat{\rho}'(t) | \mathbf{n} \rangle.
\end{equation}
All auxiliary density operators contribute to the physical state, not only the vacuum projection. The operator $\hat{\rho}'(t)$ evolves according to the pseudomode master equation (PME)
\begin{multline}\label{eq: pseudomode master equation}
    \dfrac{d\hat{\rho}'(t)}{dt}
    =
    -\dfrac{i}{\hbar} [\hat{H}_\text{eff}(t), \hat{\rho}'(t)] 
    \\
    + \sum_{j=1}^{N} \Gamma_j \big\{ 2\hat{c}_j \hat{\rho}'(t) \hat{c}^\dag_j - \hat{c}_j^\dag \hat{c}_j \hat{\rho}'(t) - \hat{\rho}'(t) \hat{c}_j^\dag \hat{c}_j \big\}.
\end{multline}
Compared with Eq. \eqref{eq: HEOM}, the dissipative part now takes the standard Lindblad form \cite{Manzano2020, lindblad1976generators, gorini1976completely}. The initial condition remains the same as in Eq. \eqref{eq: HEOM initial condition}: $\hat{\rho}'(0)=\hat{\rho}_S(0)\otimes |\mathbf{0}\rangle\langle \mathbf{0}|$. The functions $f_j(t)$ appear only in the unitary part of the master equation, specifically in $\hat{H}_\text{eff}(t)$, which is now Hermitian.

The computational complexity of solving Eq. \eqref{eq: pseudomode master equation} in Liouville space can be reduced to Hilbert-space scaling, at the cost of introducing statistical sampling through the standard Markovian quantum-state diffusion formalism \cite{Disi1998, Gisin1992, Gatarek1991}. The corresponding white-noise pseudomode stochastic Schrödinger equation (PSSE) is given in Appendix \ref{sec: PSSE}, and will be used in our numerical analysis.

\section{Benchmarking with squeezed reservoirs}\label{sec: numerical}

\begin{figure*}[t!]
    \centering
    \includegraphics[width = 0.95\linewidth]{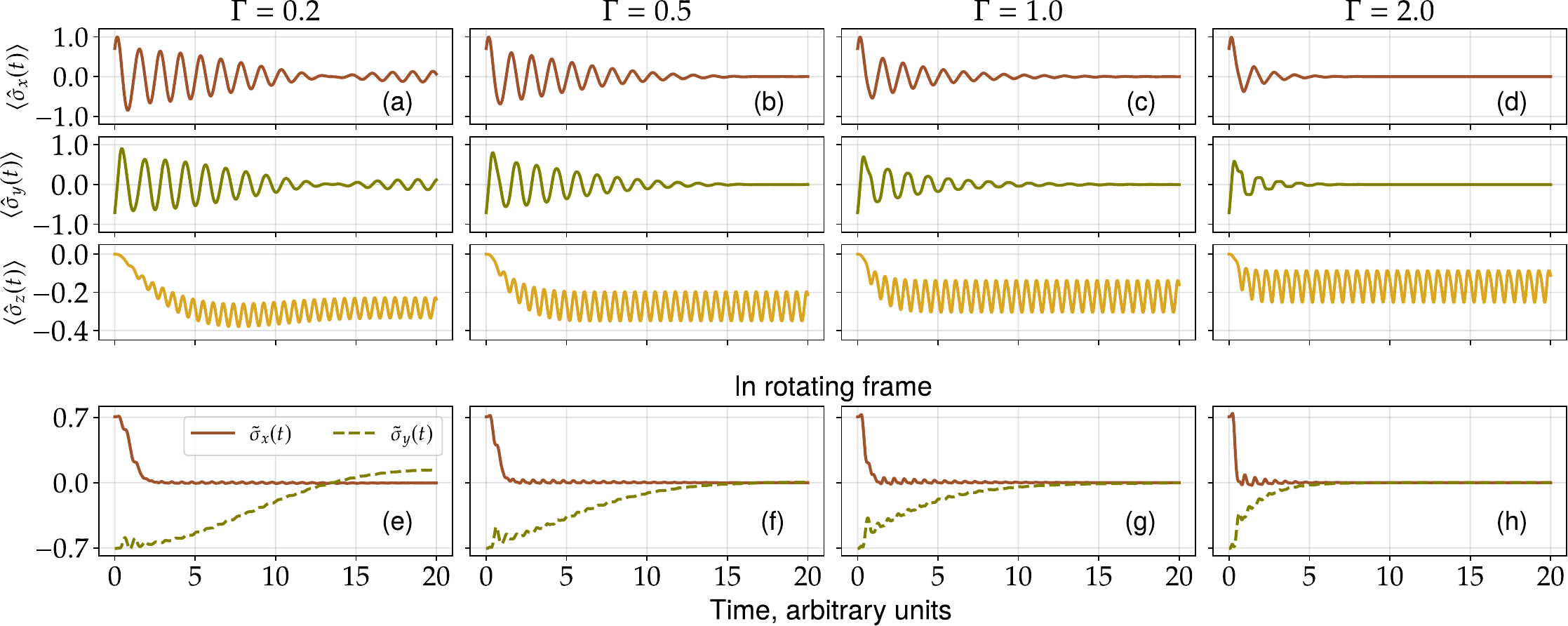}
    \caption{Expectation values of the Bloch vector components $\langle \hat{\sigma}_\alpha(t) \rangle$, where $\alpha=x,y,z$, computed for different values of $\Gamma$. The parameters are fixed as $\omega_0 = 5$, $\gamma = 1$, $r = 1.5$, and $\varphi = 0$. The atom is initially prepared in the pure state $\frac{|e\rangle + e^{-i\pi/4} |g\rangle}{\sqrt{2}}$. The dynamics is obtained by solving Eq. \eqref{eq: pseudomode master equation}, using $100$ basis states for the pseudo-Fock space. The lower row shows the Bloch vector components in the rotating frame, defined as $\tilde{\sigma}_x(t) = \langle\hat{\sigma}_+(t) \rangle e^{-i\omega_0 t} + \text{c.c.}$ and $\tilde{\sigma}_y(t) = -i\langle\hat{\sigma}_+(t) \rangle e^{-i\omega_0 t} + \text{c.c.}$}
    \label{fig: atomic dynamics}
\end{figure*}

To illustrate the applicability of the formalism in a simple yet instructive setting, we consider a two-level atom coupled to a squeezed reservoir. The squeezing is generated by a quadratic nonlinear process, which preserves the linear form of the bath coupling operator and ensures that the bath state remains Gaussian. By varying the squeezing parameters, one can access different dynamical regimes ranging from strongly non-Markovian to nearly Markovian. In this section, the parameters are chosen for computational convenience.

We consider a single two-level atom, with $|g\rangle$ and $|e\rangle$ denoting the ground and excited states. We denote the transition frequency by $\omega_0$, such that the free Hamiltonian reads
\begin{align*}
    \hat{H}_S = \dfrac{\hbar \omega_0}{2} \hat{\sigma}_z,
\end{align*}
where $\hat{\sigma}_z = |e\rangle\langle e| - |g\rangle \langle g|$. Coupling to the field is mediated by the operator $\hat{L} = \hat{\sigma}_x$. The pseudospin operators $ \hat{\sigma}_x $ and $ \hat{\sigma}_y$ can be defined using the lowering $\hat{\sigma}_- = |g\rangle\langle e|$ and raising $\hat{\sigma}_+ = |e\rangle\langle g|$ operators as:
\begin{align*}
    \hat{\sigma}_x = \hat{\sigma}_+ + \hat{\sigma}_-,
    \quad\quad
    \hat{\sigma}_y = -i\hat{\sigma}_+ + i\hat{\sigma}_-.
\end{align*}
The atom interacts with a bath of squeezed modes, for which we analyze two distinct models.

Although the derived equations of motion are formally exact, their numerical implementation requires several approximations: finite time step, truncation of the pseudo-Fock space, and, for stochastic equations, finite statistical sampling. The latter scales inversely with the square root of the number of realizations. On the other hand, truncation of the pseudo-Fock basis can introduce significant errors, particularly in the strong coupling regime or when the rotating-wave approximation cannot be applied.

In all numerical examples, the deterministic components are integrated using the fixed time-step RK4 method implemented in \texttt{DifferentialEquations.jl}~\cite{Rackauckas2017}. The master equations are solved with a step size of $T\times 10^{-5}$, where $T$ is the final simulation time. When solving stochastic equations, we take a step size of $T\times 10^{-4}$, unless otherwise stated. In the case $f_i(t)=g_i(t)$, noise terms are added following the Euler-Maruyama method at the end of each Runge-Kutta step. In the general case $f_i(t) \neq g_i(t)$, the noise is generated by diagonalizing the BCF discretized on a $10^4 \times 10^4$ grid, retaining all eigenvectors. For further details on the sampling procedure, see Appendix \ref{Sampling Z}.

\subsection{Single mode BCF}\label{sec: Strunz single mode}

\begin{figure*}[t!]
    \centering
    \includegraphics[width = \linewidth]{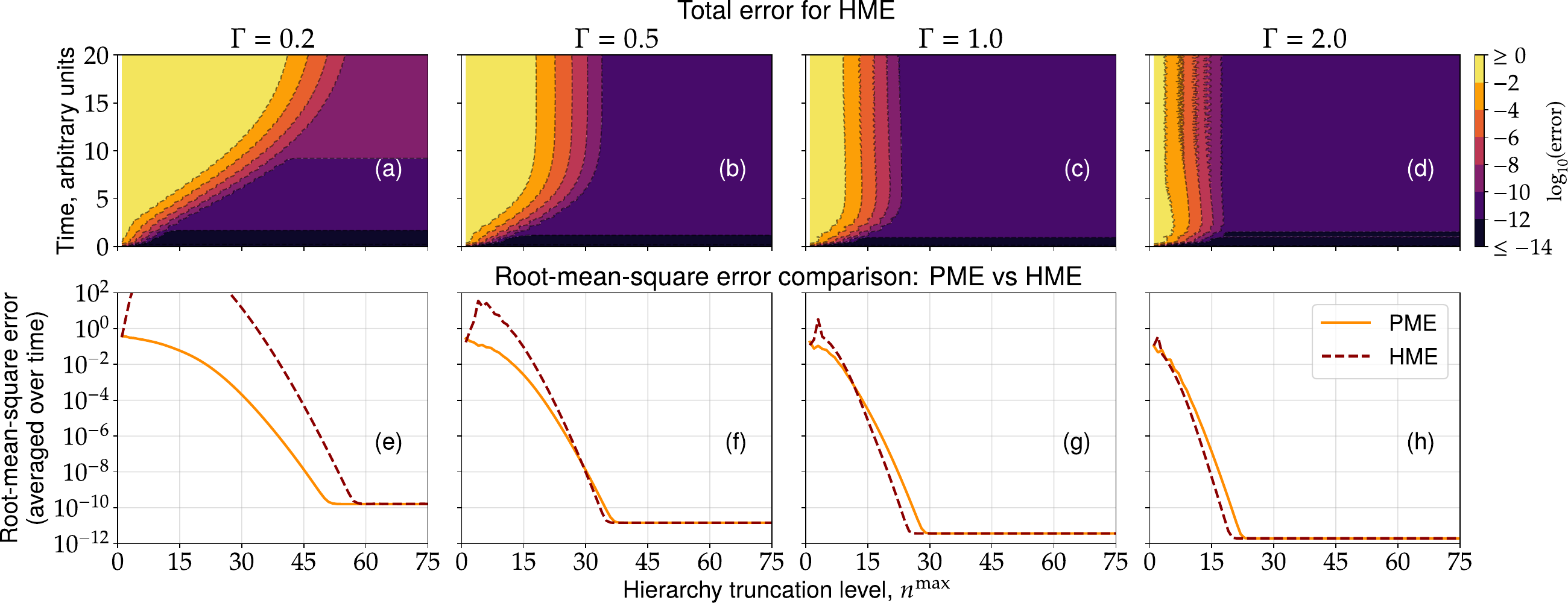}
    \caption{Comparison of the numerical error of the hierarchy of master equations (HME) \eqref{eq: HEOM} and pseudomode master equation (PME) \eqref{eq: pseudomode master equation} for different values of $\Gamma$. The other parameters are the same as in Fig. \ref{fig: atomic dynamics}: $\omega_0=5$, $\gamma=1$, $r=1.5$, and $\varphi = 0$. The total error includes contributions from the integration method and from hierarchy truncation [see Appendix \ref{sec: error ME}]. The upper row (a)-(d) shows the time dependence of the error for the hierarchy of master equations. The corresponding plot for the pseudomode master equation is not shown, as it exhibits the same qualitative behavior. The lower row (e)-(h) shows the root-mean-square error. For both approaches, the reference density operator is obtained by solving the corresponding equation (Eq.\eqref{eq: HEOM} or Eq.\eqref{eq: pseudomode master equation}) with a hierarchy depth $n^\text{max} = 100$.}
    \label{fig: ME truncation}
\end{figure*}
\begin{figure*}[t!]
    \centering
    \includegraphics[width = \linewidth]{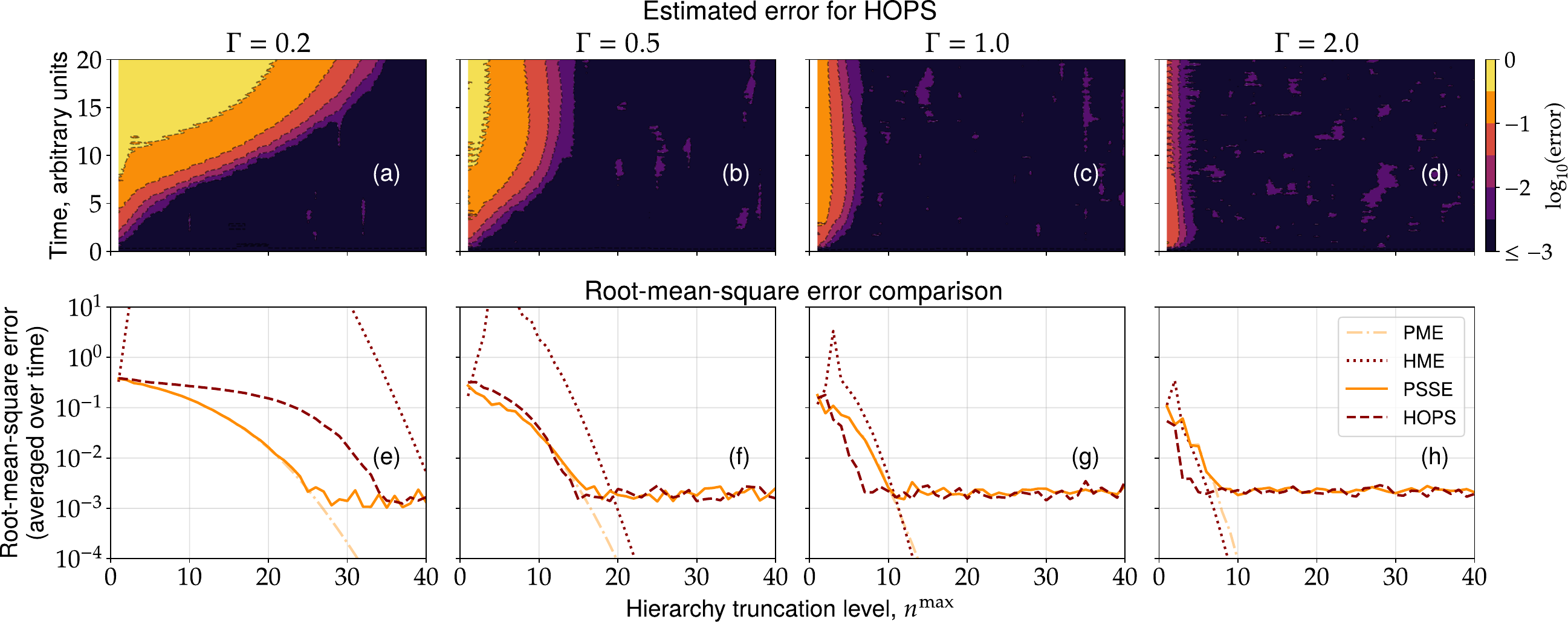}
    \caption{Estimated numerical error for the HOPS method and the pseudomode stochastic Schrödinger equation (PSSE). The other parameters are the same as in Figs. \ref{fig: atomic dynamics} and \ref{fig: ME truncation}. Statistical averages were obtained using $10^5$ trajectories. For the PSSE case with $\Gamma = 0.2$ shown in panel (e), a time step of $T \times 10^{-5}$ was used; in all other cases, the time step was $T \times 10^{-4}$. The upper row (a)-(d) shows the time dependence of the error for HOPS. The corresponding plot for the PSSE is not shown, as it exhibits the same qualitative behavior. The lower row (e)–(h) shows the time-averaged Euclidean norm of the error across different methods. These panels also include the mean errors for HME and PME from the corresponding panels in Fig. \ref{fig: ME truncation} for direct comparison. The reference density matrix was the same as in Fig. \ref{fig: ME truncation}.}
    \label{fig: SSE truncation}
\end{figure*}

\begin{subequations}\label{eq: single mode BCF}
    As a first example, we consider the BCF introduced in Ref. \cite{Link2022}. The starting point is a zero-temperature bath whose correlation function is approximated by a single exponential function. Applying a uniform squeezing operator, i.e., squeezing the whole spectrum uniformly with squeezing parameter $r$ and phase $\varphi$, yields the nonstationary correlation function
    \begin{equation}
        \alpha(t,s) = \dfrac{\Gamma}{2} e^{-\Gamma|t-s|} f(t) f^*(s),
    \end{equation}
    where the function $f(t)$ is given by
    \begin{equation}\label{eq: nonstationary function}
        f(t) = \sqrt{\gamma}  \big\{ u e^{-i(\omega_0 t - \varphi/2)} - v e^{i(\omega_0 t - \varphi/2)} \big\}.
    \end{equation}
    This form follows from the Bogoliubov transformation generated by the squeezing operator. Here, $u=\cosh{r}$ and $v=\sinh{r}$ are the Bogoliubov coefficients, satisfying $u^2-v^2 = 1$. The parameter $\gamma>0$ characterizes the system-bath coupling strength. In the unsqueezed case ($r=0$), one has $f(t) \propto \sqrt{\gamma} e^{-i\omega_0 t}$, recovering the stationary form of Eq. \eqref{eq: stationary BCF}.
\end{subequations}

For this BCF, Ref. \cite{Link2022} derived a stochastic Schrödinger equation involving two effective modes, while the numerical demonstration used the corresponding hierarchy of master equations with four modes. In contrast, our approach is more efficient: only one mode is needed for the stochastic pure-state formulation and two modes for the HME. Truncating the pseudo-Fock space to $0 \leq n \leq n^\text{max}$ results in a Hilbert space of dimension $d = 2\times (n^\text{max}+1)$ and a Liouville space of dimension $d^2$.

The atomic dynamics induced by the bath with the correlation function in Eq. \eqref{eq: single mode BCF} can also be treated using the pseudomode approach described in Sec. \ref{sec: pseudomode master equation}. Thus, this model provides an ideal test case for benchmarking and comparing the performance of all methods considered in this article. 
For the analysis, we fix the coupling strength $\gamma = 1$, central frequency $\omega_0=5$, and squeezing parameters $r=1.5$ and $\varphi=0$, while varying the spectral half-width $\Gamma$. 

Figure \ref{fig: atomic dynamics} shows the time evolution of the mean Bloch vector components for $\Gamma= 0.2$ (a), $0.5$ (b), $1.0$ (c), and $2.0$ (d). The atom evolves into a nonstationary state with oscillations in the mean $z$ component. Increasing $\Gamma$ leads to a faster decay of the transverse Bloch vector components. The bottom row (e)-(h) displays the corresponding components in the rotating frame, highlighting that the $x$ and $y$ components decay at different rates---an effect of squeezing \cite{Gardiner1986}. In this plot, we numerically solve the pseudomode master equation \eqref{eq: pseudomode master equation} with $n^\text{max} = 100$. In practice, significantly fewer hierarchies are often sufficient to reach the desired accuracy. In what follows, we show that a finite number of hierarchies is sufficient to reproduce this long-time nonstationary dynamics.

Figure \ref{fig: ME truncation} compares the numerical errors of the HME \eqref{eq: HEOM} and the PME \eqref{eq: pseudomode master equation} solved in the truncated pseudo-Fock space. The estimated error includes both the hierarchy truncation error and discretization error arising from the finite time step [see Appendix \ref{sec: error ME} for error expressions]. 

The top row (a)-(d) shows the time dependence of the error for the HME. As can be seen in panel (a), simulating longer times may require a larger number of hierarchies. In panels (b)-(d), as the parameter $\Gamma$ increases, the time after which the error remains effectively constant occurs earlier, provided that $n^\text{max}$ is sufficiently large. The time dependence of the error for the PME exhibits similar behavior and is not shown.

The bottom row (e)-(h) presents the time-averaged root-mean-square error for both methods. In panel (e), for $\Gamma = 0.2$, the PME achieves better accuracy with a much smaller truncation level. For $\Gamma = 0.5$ (f), the error for both methods becomes comparable after $30$ hierarchies. In the near-Markovian regime shown in panels (g) and (h), the HME requires fewer hierarchies than the PME. As we move closer to the non-Markovian regime in panel (e), the HME error does not decrease monotonically with increasing the number of hierarchies: it first grows and only then decreases. By contrast, the PME remains stable---adding more hierarchy levels does not increase the error. In all cases, the error curves eventually saturate, approaching a constant baseline determined by the fourth-order Runge-Kutta accuracy.

The corresponding analysis for the Girsanov‐transformed HOPS and PSSE is shown in Fig. \ref{fig: SSE truncation}. Expressions for error estimation are provided in Appendix \ref{sec: error SSE}. The top row (a)-(d) illustrates the time dependence of the error for HOPS: increasing $\Gamma$ shifts the constant‐error plateau to earlier times. The PSSE displays qualitatively similar behavior and is not shown. 

The bottom row (e)-(h) shows root-mean-square error. For $\Gamma = 0.2$ in panel (e), the PSSE achieves higher accuracy with a much smaller number of hierarchies. At $\Gamma = 0.5$ in panel (f), the error of both methods becomes comparable after moderate truncation levels. In the near‐Markovian regime in panels (g)-(h), the HOPS outperforms the PSSE. 

Unlike the HME, which demonstrated instability at small truncation levels, both stochastic approaches remain stable, with the error either decreasing or saturating at the baseline defined by Monte Carlo sampling. Panels (e)-(h) also include the mean master-equation errors from Fig. \ref{fig: ME truncation} for comparison. The PSSE error closely follows the PME error until it reaches the Monte Carlo baseline. This behavior is expected, as ensemble average of stochastic state vectors in the truncated space converges to truncated solution of the PME.

\begin{figure*}[t!]
    \centering
    \includegraphics[width = \linewidth]{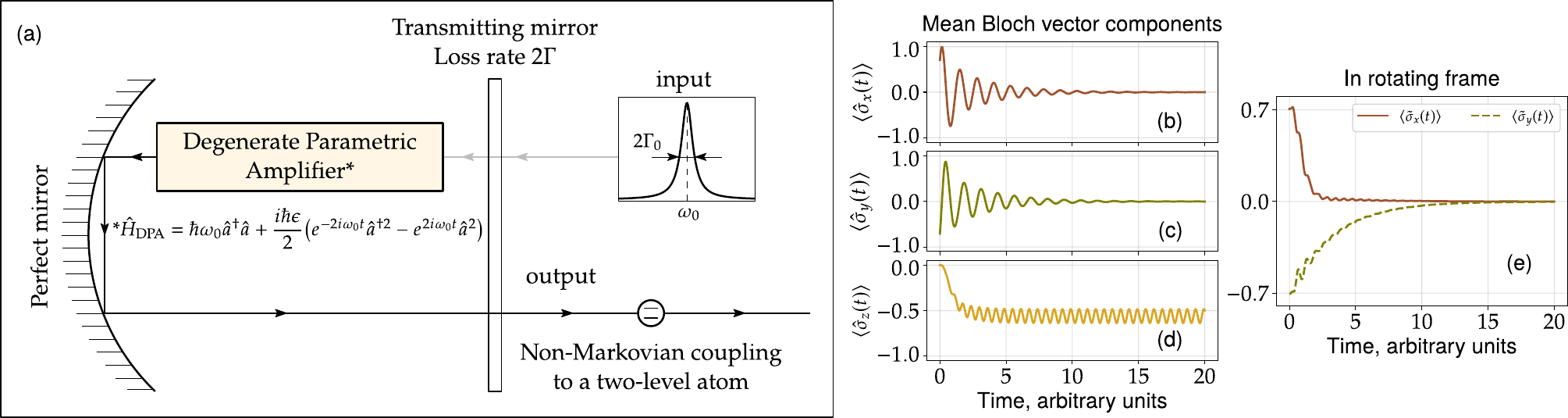}
    \caption{(a) Squeezed light generation in a one-sided cavity with a degenerate parametric amplifier. A nonlinear crystal inside the cavity is driven at frequency $2\omega_0$ with pump amplitude $\epsilon$, and the down-conversion process is described by $\hat{H}_\text{DPA}$. After reflection from the perfect mirror, the intracavity field exits through the transmitting mirror and interacts with a two-level atom, located far from the cavity. The vacuum input is modeled with a Lorentzian spectrum centered at $\omega_0$ and width $2\Gamma_0$. (b)–(e) Mean Bloch vector dynamics of the atom under this driving, characterized by the BCF in Eq. \eqref{eq: Gardiner BCF}. Parameters: $\omega_0 = 5$, $\Gamma_0=2$, $\Gamma = 1$, $\epsilon = 0.5$, $\gamma = 1$ and $\varphi = \pi$. The atom is initially in the state $\frac{|e\rangle + e^{-i\pi/4} |g\rangle}{\sqrt{2}}$.}
    \label{fig: atomic dynamics Gardiner}
\end{figure*}

In contrast, the HOPS error differs substantially from the HME error. This mismatch can be expected because, in a truncated pseudo-Fock space, the ensemble average of HOPS stochastic projectors generally does not reproduce the truncated HME density operator, $\mathbb{E} \big[ | \Psi_{\text{trunc}}(t) \rangle \langle \Psi_{\text{trunc}}(t) | \big] \not= \hat{\rho}_{\text{trunc}}(t)$,
since the derivation of the HME \eqref{eq: HEOM} relies on retaining the full, untruncated basis. The stability of HOPS---and its higher tolerance to hierarchy truncation compared with the HME---can be attributed to the fact that part of the bath influence is carried by the stochastic term $Z(t)$, in addition to the hierarchy itself. As a result, truncating the pseudo-Fock space has a weaker impact, because the statistical properties of $Z(t)$ do not depend on the truncation. We emphasize that our HME benchmark was performed only for the specific bath-correlation function considered here; therefore, we cannot draw general conclusions about the stability of the HME.

\subsection{Three mode BCF}

\begin{subequations}\label{eq: Gardiner BCF}
    A different model of a squeezed-light reservoir was considered in Refs. \cite{Gardiner1984, Collett1984}. It describes squeezed light generated by a degenerate parametric amplifier in a one-sided cavity with one perfectly reflecting and one partially transmitting mirror. The cavity loss rate is denoted by $2\Gamma$. Input vacuum fluctuations enter through the transmitting mirror and interact with a continuously pumped nonlinear crystal characterized by an effective coupling constant $\epsilon > 0$. After reflection from the perfect mirror, the generated field exits through the transmitting mirror and drives a two-level atom [see Fig. \ref{fig: atomic dynamics Gardiner} (a)]. The resulting non-Markovian dynamics can be treated with the hierarchy of pure states, which requires only the bath correlation function of the output field.
    
    We assume that the input vacuum has a Lorentzian spectral density with FWHM $2\Gamma_0$. The corresponding input BCF is
    \begin{equation*}
    \alpha_\text{input}(t-s) = \dfrac{\gamma \Gamma_0}{2} \, e^{-\Gamma_0 |t-s| -i\omega_0(t-s)},
    \end{equation*} 
    where $\gamma$ describes the atom-field coupling. Introducing this finite bandwidth serves to regularize the correlation function of the input quantum white noise. We then formulate the input–output relations under the assumption $\Gamma \ll \Gamma_0 \ll \omega_0$, corresponding to the Markovian regime of cavity leakage. In the limit $\Gamma_0 \rightarrow \infty$, the spectral density becomes flat, corresponding to quantum white noise.
    
    Using the input-output formalism \cite{Collett1984, Gardiner1985}, we find the correlation function for the output field at the position of the atom
    \begin{equation}
        \alpha(t,s) = \sum_{i=1}^{3} \alpha_i(t-s) f_i(t) g_i^*(s),
    \end{equation}
    which is composed of three effective modes with decay rates $\Gamma_1 = \Gamma_0$, $\Gamma_2 = \Gamma - \epsilon$, and $\Gamma_3 = \Gamma + \epsilon$. Here, $\alpha_j(\tau)$ are exponential kernels [see Eq. \eqref{eq: exponential kernel}]. The nonstationary functions $f_i(t)$ are given by
    \begin{align}
        &f_1(t) = \sqrt{\gamma} \big\{ u e^{-i(\omega_0t - \varphi/2)} - v e^{i(\omega_0t - \varphi/2)} \big\},\\
        &f_2(t) = \sqrt{\dfrac{4\gamma \Gamma \epsilon}{\Gamma_-^2} \dfrac{\Gamma_0^2}{\Gamma_0^2-\Gamma_-^2}}\cos(\omega_0 t - \varphi/2),\\
        &f_3(t) = \sqrt{\dfrac{4\gamma \Gamma \epsilon}{\Gamma^2_+} \dfrac{\Gamma_0^2}{\Gamma_0^2-\Gamma_+^2}}\sin(\omega_0 t - \varphi/2),
    \end{align}
    where $\Gamma_\pm = \Gamma \pm \epsilon$ and $\varphi$ is the phase of the field at the atom. For the functions $g_j(t)$, we have 
    \begin{equation}
    g_1(t) = f_1(t), 
    \quad g_2(t) = f_2(t), 
    \quad g_3(t) = - f_3(t).
    \end{equation}
    Because of the sign flip in the third mode, the pseudomode representation is inapplicable. Equation \eqref{eq: Gardiner BCF} is valid for $\epsilon < \Gamma$ (i.e., below threshold) and $\Gamma_0 > \Gamma_\pm$. In the limit $\Gamma_0 \rightarrow \infty$, the first mode reduces to a delta function, and the BCF simplifies to the form studied in Refs. \cite{Parkins1988, Collett1984, Ritsch1988}. If additionally both $\Gamma_\pm$ are large, all exponential kernels approach delta functions, giving rise to squeezed white noise \cite{Gardiner1986}.

\begin{figure*}[t!]
    \centering
    \includegraphics[width = \linewidth]{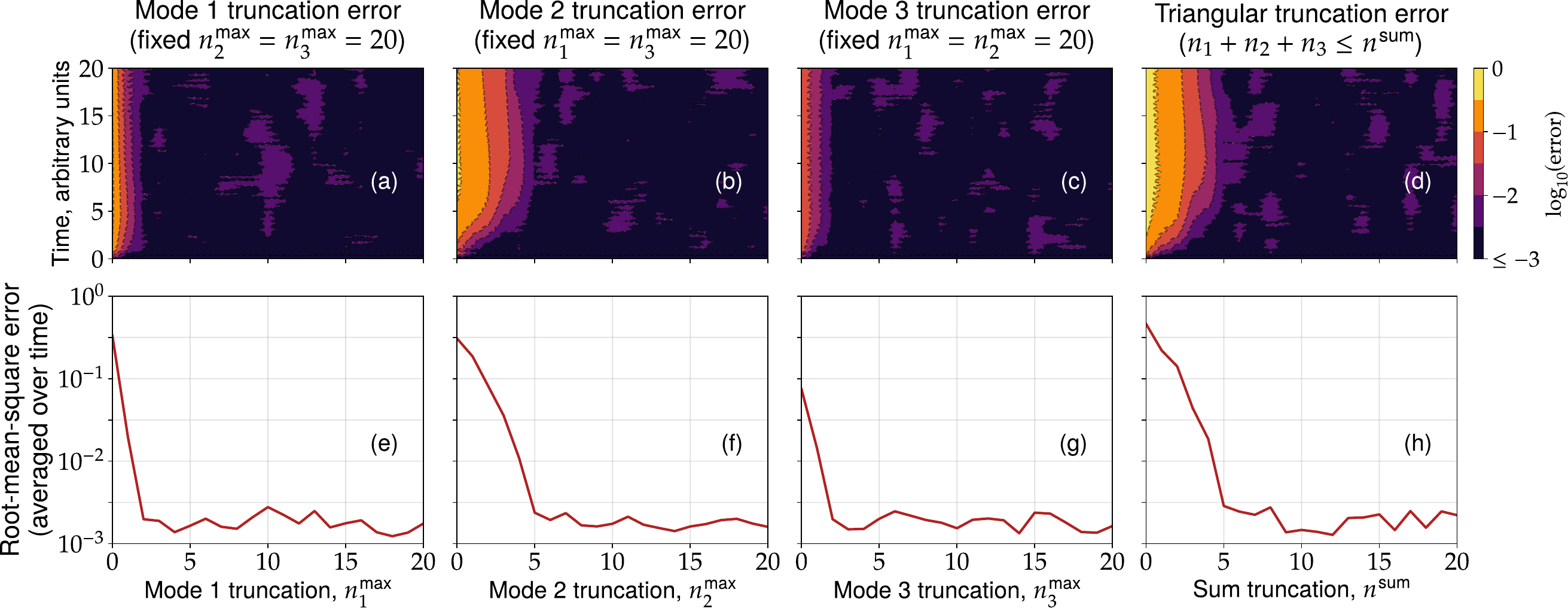}
    \caption{Estimated numerical error of the HOPS method. Parameters are the same as in Fig. \ref{fig: atomic dynamics Gardiner}. Statistical averages were obtained from $10^5$ trajectories with a time step of $20\times10^{-4}$. The upper row (a)–(d) shows the time dependence of the error: (a) $n^{\text{max}}_2=n^{\text{max}}_3=20$, varying $n^{\text{max}}_1$; (b) $n^{\text{max}}_1=n^{\text{max}}_3=20$, varying $n^{\text{max}}_2$; (c) $n^{\text{max}}_1=n^{\text{max}}_2=20$, varying $n^{\text{max}}_3$; (d) triangular truncation with $n_1+n_2+n_3\leq n^{\text{sum}}$, varying $n^{\text{sum}}$. The lower row (e)–(h) shows the corresponding time-averaged errors. The reference density matrix is obtained using the same time step, averaging over $10^6$ trajectories and truncation levels $n^{\text{max}}_1=n^{\text{max}}_2=n^{\text{max}}_3=20$.}
    \label{fig: SSE truncation Gardiner}
\end{figure*}

    The first mode has the same structure as the single-mode BCF in Eq. \eqref{eq: single mode BCF}, but with Bogoliubov coefficients
    \begin{align*}
        &u = \dfrac{\Gamma_0^{2} - \Gamma^2 - \epsilon^2}{\sqrt{(\Gamma_0^{2} - \Gamma_+^{2})(\Gamma_0^{2} - \Gamma_-^{2})}},
        & v = \dfrac{2\Gamma\epsilon}{\sqrt{(\Gamma_0^{2} - \Gamma_+^{2})(\Gamma_0^{2} - \Gamma_-^{2})}},
    \end{align*}
    which are constrained by the physical parameters. In the limit $\Gamma_0\rightarrow \infty$, the squeezing of this mode disappears, as $u\rightarrow 1$ and $v \rightarrow 0$. 
    The second mode has decay rate $\Gamma_2=\Gamma_-$, which decreases as $\epsilon \rightarrow \Gamma$. The corresponding effective coupling strength ($\propto |f_2|^2$) scales as $\Gamma_-^{-1}$ and becomes large near threshold. The third mode, with decay rate $\Gamma_3 = \Gamma_+$, enters the BCF with a negative sign. The effective squeezing parameter can be estimated as
    \begin{equation*}
        \tilde{r} = \text{arccosh} \frac{\Gamma}{\sqrt{\Gamma^2 - \epsilon^2}}.
    \end{equation*}
    Ideal squeezing is achieved at the threshold, when $\epsilon \approx \Gamma$ \cite{Ritsch1988, Parkins1988}.

    The characteristic atomic dynamics is shown in Fig. \ref{fig: atomic dynamics Gardiner} (b)-(e): panels (b)-(d) display the mean Bloch vector components, while panel (e) shows the mean transverse components in the rotating frame. We use the parameters $\omega_0 = 5$, $\Gamma_0=2$, $\Gamma = 1$, $\epsilon = 0.5$, $\gamma = 1$ and $\varphi = \pi$, which correspond to the effective squeezing $\tilde{r} \approx 0.55$. Each mode is truncated at $20$ hierarchies.

    Let $n_j^{\text{max}}$ denote the truncation level of the $j$th mode, with $0\leq n_j \leq n_j^{\text{max}}$. The dimension of the Hilbert space of the atom plus three truncated modes is
    \begin{equation*}
        d = 2 \times \prod_{j=1}^{3} (n_j^{\text{max}} + 1),
    \end{equation*}
    and the corresponding Liouville space has a dimension of $d^2$. This rapid scaling limits direct use of the HME. For example, with $n_j^{\text{max}}= 9$ ($10$ pseudo-Fock states per mode), one obtains $d^2 = 4\times 10^6$, whereas HOPS requires only $d = 2\times 10^3$ basis states.

    To assess truncation errors [see Appendix \ref{sec: error SSE} for expressions], Fig.~\ref{fig: SSE truncation Gardiner} (a)–(c) shows the error when varying the truncation level of one mode while fixing the others. All modes quickly reach a time-independent error plateau. Panels (e)–(g) display the corresponding root-mean-square errors, indicating that modes 1 and 3 are nearly Markovian, while mode 2 requires the largest number of hierarchies. Thus, taking $n_j^{\text{max}} = 20$ for all modes is unnecessary at the given Monte Carlo baseline; only the second mode needs many hierarchies.

    Another strategy is triangular truncation, where instead of bounding each $n_j$ individually, one restricts their sum: $n_1+n_2+n_3 \leq n^\text{sum}$. This changes the Hilbert space dimension to
    \begin{equation*}
        d_\text{triang} = 2 \times \binom{n^\text{sum}+3}{n^\text{sum}}.
    \end{equation*}
    Panels (d) and (h) of Fig.~\ref{fig: SSE truncation Gardiner} demonstrate that this approach is quite effective and illustrative: the values of $n^\text{sum} \simeq 5$ are sufficient to reach a Monte Carlo baseline.
\end{subequations}

\section{Discussion and Conclusion}\label{sec: discussion}

In summary, we generalized the hierarchy of pure states method to open quantum systems linearly coupled to nonstationary Gaussian baths by introducing a nonstationary decomposition of the BCF (Sec. \ref{sec: OQS}). This maintains the favorable scaling of HOPS while extending its applicability beyond stationary baths, requiring only minor changes to the noise sampling and effective Hamiltonian. Benchmarks with squeezed reservoirs in Sec. \ref{sec: numerical} showed rapid convergence with respect to the hierarchy depth. 
Our approach assumes only a Gaussian bath and linear coupling, therefore, its scope extends beyond the benchmarks studied here. Below we show how to apply it to a broader class of Gaussian baths and how to incorporate finite temperature.

\subsection{Displaced squeezed thermal state}

A general Gaussian bath state can be generated by displacing and squeezing a suitable thermal state $\hat{\rho}_\text{th}$ \cite{Adam1995}:
\begin{equation}\label{eq: displaced squeezed thermal state}
    \hat{\rho}_B = \hat{D}(\bm{\alpha}) \hat{S}  \hat{\rho}_\text{th}  \hat{S}^\dag \hat{D}^\dag(\bm{\alpha}),
\end{equation}
where $\hat{D}(\bm{\alpha})$ is the displacement operator and $\hat{S}$ is a general squeezing operator. The bath is linearly coupled to the system, with the Hamiltonian in Eq. \eqref{eq: Hamiltonian general}, and the bath operator
\begin{equation}\label{eq: bath operator original}
    \hat{B}(t) = \sum_\lambda g_\lambda e^{-i\omega_\lambda t} \hat{b}_\lambda + \text{H.c.}
\end{equation}
We anticipate that a hierarchy of pure states can be derived for any such bath using our results, provided the effective BCF is represented as in Eq. \eqref{eq: BCF ansatz}.

A drawback of constructing the hierarchy directly for the state \eqref{eq: displaced squeezed thermal state} is that it populates auxiliary states at $t=0$, making the hierarchy explicitly dependent on the temperature, squeezing, and displacement parameters. To avoid this, our derivation in Appendix \ref{sec: Appendix A} introduces an effective bath in the vacuum state together with a modified bath operator that reproduces the BCF. Here we illustrate the procedure for Eq. \eqref{eq: displaced squeezed thermal state}. 

Following Refs. \cite{Disi1998, Hartmann2017}, we map the finite-temperature bath to an effective zero-temperature one. Temperature enters through an average over classical displacements, which is equivalent to adding a stochastic contribution to the Hamiltonian. Using the Glauber-Sudarshan representation, the thermal state is written as a Gaussian mixture of coherent states:
\begin{equation}\label{eq: P representation thermal}
    \hat{\rho}_\text{th} = \int \! \prod_\lambda \dfrac{d^2 y_\lambda}{\pi \overline{n}_\lambda} e^{-|y_\lambda|^2/\overline{n}_\lambda} \, \hat{D}(\mathbf{y}) |0\rangle \langle 0 | \hat{D}^\dag(\mathbf{y}),
\end{equation}
with $\overline{n}_\lambda = (e^{\hbar\omega_\lambda / k_B T} - 1)^{-1}$ and $\hat{D}(\bm{y}) = \prod_\lambda \exp(y_\lambda \hat{b}_\lambda^\dag - y^*_\lambda \hat{b}_\lambda)$. Substituting this into Eq. \eqref{eq: displaced squeezed thermal state} reveals that this state is obtained by first applying the unitary
\begin{equation}
    \hat{U} = \hat{D}(\bm{\alpha}) \hat{S} \hat{D}(\bm{y})
\end{equation}
to the vacuum, so that $\hat{U}|0\rangle\langle 0| \hat{U}^\dag$ is the state for a given $\bm{y}$, and then averaging over $\bm{y}$ with the Gaussian weight as in Eq. \eqref{eq: P representation thermal}. 

To find the reduced system dynamics, it is convenient to absorb $\hat{U}$ into the interaction Hamiltonian. This yields the unitary transformation of the bath operator,
\begin{equation}\label{eq: bath operator transformation}
    \hat{U}^\dag \hat{B}(t)  \hat{U} = \mathcal{B}(t) + \hat{B}_\text{eff}(t) + Y(t),
\end{equation}
with the bath now in the vacuum state. The reduced density matrix is then obtained by evolving with the modified interaction Hamiltonian and averaging over $\bm{y}$ with the Gaussian weight of Eq.~\eqref{eq: P representation thermal}.

Each term in Eq. \eqref{eq: bath operator transformation} has a clear origin in $\hat{U}$. The displacement operator $\hat{D}(\bm{\alpha}) = \prod_\lambda \exp(\alpha_\lambda \hat{b}_\lambda^\dag - \alpha^*_\lambda \hat{b}_\lambda)$ contributes a classical driving field $\mathcal{B}(t)$,
\begin{equation*}
    \mathcal{B}(t) = \sum_\lambda g_\lambda e^{-i\omega_\lambda t} \alpha_\lambda + \text{c.c.}
\end{equation*}
Squeezing $\hat{S}$ mixes the creation and annihilation operators via a Bogoliubov transformation, which dresses the bath operator
\begin{equation}\label{eq: effective bath operator}
   \hat{B}_\text{eff}(t) = \hat{S}^\dag \hat{B}(t) \hat{S} = \sum_\lambda g_\lambda(t) \hat{b}_\lambda + \text{H.c.},
\end{equation}
where the explicit form of $g_\lambda(t)$ depends on the squeezing model. Finally, applying the displacement $\hat{D}(\bm{y})$ to the dressed bath operator $ \hat{B}_\text{eff}(t)$ yields the additional term $Y(t)$ that describes finite-temperature fluctuations
\begin{equation}
   Y(t) = \sum_\lambda g_\lambda(t) y_\lambda + \text{c.c.}
\end{equation}
Interpreting $\bm{y}$ as Gaussian random variables distributed according to the weight in Eq. \eqref{eq: P representation thermal}, $Y(t)$ becomes a real-valued stochastic process with zero mean and nonstationary two-time correlation function
\begin{equation}
    \mathbb{E} \big[ Y(t) Y(s) \big] = \sum_\lambda 2\overline{n}_\lambda \mathrm{Re} [g_\lambda(t) g_\lambda^*(s)].
\end{equation}
Sampling this noise is a separate topic and will not be covered here. This approach eliminates the explicit dependence of the hierarchy on temperature. As demonstrated in Ref. \cite{Hartmann2017}, it shows better performance when truncating the hierarchy compared to encoding temperature in the initial condition.

Coupling to the effective bath in the vacuum state is mediated by the bath operator in Eq. \eqref{eq: effective bath operator}. Consequently, the bath correlation function has the simplified, temperature-independent form
\begin{equation}\label{eq: effective BCF}
    \alpha(t,s) = \langle 0 | \hat{B}_\text{eff}(t)  \hat{B}_\text{eff}(s) | 0 \rangle = \sum_\lambda g_\lambda(t) g^*_\lambda(s),
\end{equation}
which is essentially determined by the squeezing model. Once $\alpha(t,s)$ is fitted to the ansatz in Eq. \eqref{eq: BCF ansatz}, the stochastic Schrödinger equation retains the form of Eq. \eqref{eq: SSE (I)} while the effective Hamiltonian acquires a deterministic drive $\mathcal{B}(t)$ and a thermal-noise term $Y(t)$:
\begin{equation*}
    \hat{H}_\text{eff}(t) \rightarrow \hat{H}_\text{eff}(t) + \hbar \{ \mathcal{B}(t)+ Y(t) \} \hat{L}.
\end{equation*}
Finally, the reduced density matrix and observables are obtained by performing an additional average over realizations of $Y(t)$ in Eqs. \eqref{eq: reduced density matrix sampling} and \eqref{eq: expectation values}.

\subsection{Uniform squeezing model}

As a concrete example from the literature, consider the uniform two-mode squeezing operator $\hat{S} = \hat{S}(r,\varphi)$ with squeezing parameter $r>0$ and phase $\varphi$:
\begin{equation}
    \hat{S}(r,\varphi) = \prod_\lambda \exp \Big\{ \dfrac{r}{2} \big( e^{-i\varphi} \hat{b}_{2\lambda_0-\lambda} \hat{b}_\lambda - e^{i\varphi} \hat{b}^\dag_\lambda \hat{b}^\dag_{2\lambda_0-\lambda} \big) \Big\}.
\end{equation}
It pairs modes symmetrically around the central frequency $\omega_0$ (indexed by $\lambda_0$). It induces time-dependent effective couplings
\begin{equation}
   g_\lambda(t) = g_\lambda S(t) e^{-i\omega_\lambda t},
\end{equation}
where $S(t) = u - v e^{i(2\omega_0t-\varphi)}$, with $u=\cosh{r}$ and $v=\sinh{r}$. In deriving this, we assume real couplings $g_\lambda$ and mirror symmetry $g_\lambda = g_{2\lambda_0-\lambda}$. The BCF \eqref{eq: effective BCF} then becomes
\begin{equation}
    \alpha(t,s) = S(t) S^*(s)  \sum_\lambda g_\lambda^2 e^{-i\omega_\lambda(t-s)}.
\end{equation}
Thus, it is the unsqueezed (vacuum) correlation function dressed by the squeezing transformation. Approximating the unsqueezed BCF by a sum of exponentials, as in Eq. \eqref{eq: stationary BCF},
\begin{equation*}
    \sum_\lambda g_\lambda^2 e^{-i\omega_\lambda\tau} 
    \approx \sum_{j=1}^{N} \gamma_j \alpha_j(\tau) e^{-i\omega_j\tau},
\end{equation*}
where each $\alpha_j(\tau)$ is an exponential kernel [Eq. \eqref{eq: exponential kernel}], the resulting BCF takes the nonstationary ansatz form of Eq. \eqref{eq: BCF ansatz} with
\begin{align*}
    f_j(t) & = \sqrt{\gamma_j} \big\{ u e^{-i(\omega_0 t - \varphi/2)} - v e^{i(\omega_0 t - \varphi/2)} \big\} e^{-i\Delta_j t}, \\
    g_j(s) & = \sqrt{\gamma^*_j}  \big\{ u e^{-i(\omega_0 s - \varphi/2)} - v e^{i(\omega_0 s - \varphi/2)} \big\} e^{-i\Delta_j s},
\end{align*}
where $\Delta_j = \omega_j-\omega_0$. These functions share the structure of the single-mode expansion in Eq. \eqref{eq: nonstationary function}, augmented by phase factors oscillating at the detunings $\Delta_j$. In general, $\gamma_j$ are complex, and $f_j(t) \not = g_j(t)$. When all $\gamma_j$ are real and positive, the pseudomode representation applies, and each effective mode can be associated with an independent bath. Such squeezed baths were analyzed in Refs. \cite{Ablimit2023, Xie2025}. This setting can benefit from keeping the number of effective modes minimal and from the pseudomode representation, which becomes more effective in strongly non-Markovian regimes. For squeezed thermal reservoirs in the high-temperature limit, see Ref. \cite{Ablimit2025}.

Beyond applications in general environments out of equilibrium, we anticipate that our methods will find applications in light-matter interactions. 
For instance, spectroscopy with quantum light~\cite{Dorfman2016}, especially in regimes with large numbers of photons~\cite{Carlos2021, Panahiyan2023, schlawin2025theoryquantumenhancedstimulatedraman}, where a direct description of the light field becomes prohibitively expensive, could be described by our methods. The same holds true for driven cavity quantum materials~\cite{cavityreview}, where strong light-matter coupling enhances the influence of photonic fluctuations.

\begin{acknowledgments}
    The authors are grateful to Prof. Frank Großmann for insightful discussions and helpful comments. V.S. and F.S. acknowledge the financial support of the Cluster of Excellence ``CUI: Advanced Imaging of Matter'' of the Deutsche Forschungsgemeinschaft (DFG) --- EXC 2056 --- project ID 390715994. F.S. acknowledges support from the DFG research unit 'FOR5750: OPTIMAL' - project ID 531215165. 
    This research was supported in part through the Maxwell computational resources operated at Deutsches Elektronen-Synchrotron DESY, Hamburg, Germany.
\end{acknowledgments}

\appendix

\section{Derivation of stochastic Schrödinger equation \eqref{eq: SSE (I)}}\label{sec: Appendix A}

As already mentioned, any pair of a linear $\hat{B}(t)$ and a Gaussian initial state reproducing the BCF results in the same reduced system dynamics. This freedom allows us to replace the bath with an effective one in the vacuum state, coupled to the system via a modified operator $\hat{B}_\text{eff}(t)$ satisfying
\begin{equation}\label{eq: BCF definition}
    \langle 0 | \hat{B}_\text{eff}(t) | 0 \rangle = 0, 
    \quad\quad
    \alpha(t,s) = \langle 0 | \hat{B}_\text{eff}(t) \hat{B}_\text{eff}(s) | 0 \rangle.
\end{equation}
Choosing the vacuum state eliminates the initial conditions for effective bath modes and simplifies the subsequent derivation.

We adopt the following ansatz for $\hat{B}_\text{eff}(t)$:
\begin{equation}\label{eq: coupling modified}
\hat{B}_\text{eff}(t) = \sum_\lambda g_\lambda(t) \hat{b}_\lambda + \text{H.c.},
\end{equation}
with bosonic operators $[\hat{b}_\lambda, \hat{b}^\dagger_{\lambda'}] = \delta_{\lambda\lambda'}$ and possibly continuous mode index $\lambda$. Functions $g_\lambda(t)$ reproduce the BCF:
\begin{equation}
\alpha(t,s) = \sum_\lambda g_\lambda(t) g^*_\lambda(s).
\end{equation}
The existence of these functions follows from the positive semi-definiteness of the BCF. In what follows, we assume that the BCF can be approximated by the form \eqref{eq: BCF ansatz}.

With this effective bath in place, Section \ref{sec: A.2} derives the non-Markovian stochastic Schrödinger equation via coherent-state unraveling \cite{Disi1997}, and Section \ref{sec: A.3} introduces the corresponding hierarchy of pure states.

\subsection{Tracing out the bath in a coherent-state representation}\label{sec: A.2}

We focus on the dynamics of the system and trace out the bath degrees of freedom in the coherent-state basis. For simplicity, we assume the system is initially in a pure state. The initial system-bath density matrix must be factorized as $\hat{\rho}_S(0) \otimes \hat{\rho}_B$.

The bath is traced out using Bargmann coherent states, defined as $|\bm{z}\rangle = \prod_\lambda e^{z_\lambda \hat{b}^\dag_\lambda} |\text{vac}\rangle$, where $\bm{z}$ denotes the vector of all $z_\lambda$. This leads to the reduced density matrix of the system:
\begin{multline}
    \hat{\rho}_S(t) = \mathrm{Tr}_B \big[ |\psi_{S+B}(t) \rangle \langle \psi_{S+B}(t) | \big] 
    \\= \int \! \prod_\lambda \dfrac{d^2 z_\lambda \, e^{-|z_\lambda|^2}}{\pi}  |\psi(\bm{z}^*, t)  \rangle \langle \psi(\bm{z}^*, t) |,
\end{multline}
where $|\psi_{S+B}(t) \rangle$ is the total system-bath state. The conditional state, defined as $|\psi(\bm{z}^*, t) \rangle = \langle \bm{z} | \psi_{S+B}(t) \rangle$, depends parametrically on the complex conjugates $\bm{z}^*$ and satisfies the Schrödinger equation:
\begin{multline}\label{eq: Schrodinger equation conditional states}
    \dfrac{\partial|\psi(\bm{z}^*, t)\rangle}{\partial t} = \Big[ -\dfrac{i}{\hbar} \hat{H}_S(t) - i \mathcal{D}(t) \hat{L} \\- i Z^*(t) \hat{L} \Big] |\psi(\bm{z}^*, t)\rangle.
\end{multline}
Here, $Z(t)$ and the functional derivative $\mathcal{D}(t)$ are given by:
\begin{align}
    & Z(t) = \sum_\lambda g_\lambda(t) z_\lambda,
    & \mathcal{D}(t) = \sum_\lambda g_\lambda(t) \dfrac{\delta}{\delta z_\lambda^*}.
\end{align}
Alternatively, $\mathcal{D}(t)$ can be written in the time domain as a functional derivative with respect to $Z^*(t)$:
\begin{equation}\label{eq: functional derivative original}
    \mathcal{D}(t) = \int_{-\infty}^{+\infty} \! ds \, \alpha(t,s) \dfrac{\delta}{\delta Z^*(s)}.
\end{equation}
Since the bath is initially in the vacuum state, causality requires that the upper limit of the integral be restricted to $t$ when acting on the state at time $t$ \cite{Suess2014}.

Let $\hat{O}$ denote an operator acting on the system; its expectation value is given by
\begin{multline}\label{eq: observable conditional}
    \mathrm{Tr} [ \hat{O} \hat{\rho}_S(t) ] \\ = \int \! \prod_\lambda \dfrac{d^2 z_\lambda \, e^{-|z_\lambda|^2}}{\pi} \langle \psi(\bm{z}^*, t)| \hat{O} | \psi(\bm{z}^*, t) \rangle.
\end{multline}
This integral allows a statistical interpretation: the exponential defines a Gaussian distribution for the complex variables $z_\lambda$. However, due to the functional derivative $\mathcal{D}(t)$ in Eq. \eqref{eq: Schrodinger equation conditional states}, the time evolution of $|\psi(\bm{z}^*,t)\rangle$ is not independent across different realizations of $\bm{z}^*$, which precludes the use of standard Monte Carlo methods. The hierarchy of pure states formalism resolves this issue by eliminating the functional derivative from the Schrödinger equation.

\subsection{Eliminating the functional derivative: a hierarchy}\label{sec: A.3}

HOPS expresses the action of the functional derivative on $|\psi(\bm{z}^*, t)\rangle$ through a hierarchy of auxiliary states, thereby eliminating it from the equations. Substituting the BCF decomposition \eqref{eq: BCF ansatz} into the definition \eqref{eq: functional derivative original} yields
\begin{subequations}
    \begin{align}
        \mathcal{D}(t) & = \sum_{j=1}^{N} f_j(t) \mathcal{D}_j(t),\\
        \mathcal{D}_j(t) & = \! \int_{-\infty}^{+\infty} \! ds \, \alpha_j(t-s) g_j^*(s) \dfrac{\delta}{\delta Z^*(s)}.
    \end{align}
\end{subequations}
Each $\mathcal{D}_j(t)$ depends on $t$ only through the stationary kernel $\alpha_j(t-s)$. Consequently, their time derivatives satisfy:
\begin{equation}\label{eq: functional derivative time derivative}
    \dfrac{\partial\mathcal{D}_j(t)}{\partial t} |\psi(\bm{z}^*, t)\rangle = - \Gamma_j \mathcal{D}_j(t)  |\psi(\bm{z}^*, t)\rangle.
\end{equation}
Note that the derivative of the full operator $\mathcal{D}(t)$ cannot be expressed solely in terms of $\mathcal{D}(t)$. Thus, constructing the hierarchy from $\mathcal{D}(t)$ does not yield a closed set of equations, thus, we use instead the components $\mathcal{D}_j(t)$.

The auxiliary states are indexed by the number of times each derivative $\mathcal{D}_j(t)$ is applied, denoted by $\mathbf{n}=(n_1, \ldots, n_N)$. We interpret $\mathbf{n}$ as occupation numbers in a pseudo-Fock space \cite{Gao2022}, and define an extended state vector in the tensor product of the system's Hilbert space and the pseudo-Fock space:
\begin{equation}\label{eq: extended vector definition}
    |\Psi(\bm{z}^*, t) \rangle 
    = \mathrm{exp} \Big[ \sum_{j=1}^{N} {\sqrt{\dfrac{2}{ \Gamma_j}}\mathcal{D}_j(t) \hat{c}_j^{\dagger}} \Big] |\psi(\bm{z}^*, t)\rangle \otimes |\mathbf{0}\rangle,
\end{equation}
where $\hat{c}_j, \hat{c}^\dag_j$ are bosonic operators for the effective modes, satisfying $[\hat{c}_i, \hat{c}_j^\dag] = \delta_{ij}$. The auxiliary states are obtained as projections $|\psi^{(\mathbf{n})}(\bm{z}^*, t)\rangle = \langle \mathbf{n} |\Psi(\bm{z}^*, t) \rangle$, and the projection onto the vacuum yields a physical state $|\psi(\bm{z}^*, t)\rangle = \langle \mathbf{0} |\Psi(\bm{z}^*, t) \rangle$.

The action of the functional derivative $\mathcal{D}_j(t)$ on the extended state vector $|\Psi(\bm{z}^*, t) \rangle$ reduces to the annihilation of a quantum in the corresponding effective mode:
\begin{equation}
    \mathcal{D}_j(t)|\Psi(\bm{z}^*, t)\rangle = \sqrt{\frac{\Gamma_j}{2}} \hat{c}_j |\Psi(\bm{z}^*, t)\rangle.
\end{equation}
Using this identity together with Eq. \eqref{eq: functional derivative time derivative}, we obtain the equation of motion for the extended state vector:
\begin{multline}\label{eq: schrodinger equation extended vector}
   \dfrac{\partial |\Psi(\bm{z}^*, t)\rangle}{\partial t} = \Big[ - \sum_{j=1}^{N} \Gamma_j \hat{c}_j^\dag \hat{c}_j
   -
   \dfrac{i}{\hbar} \hat{H}_{\text{eff}}(t) 
   \\
    - i Z^*(t) \hat{L} \Big] |\Psi(\bm{z}^*, t)\rangle,
\end{multline}
where the effective Hamiltonian is defined in Eq. \eqref{eq: effective Hamiltonian}. 
Since Eq. \eqref{eq: schrodinger equation extended vector} contains no functional derivatives, a Monte Carlo implementation becomes feasible by interpreting each $z_\lambda$ as a complex Gaussian random variable satisfying
\[
\mathbb{E}[z_\lambda z^*_{\lambda'}] = \delta_{\lambda\lambda'}, \quad
\mathbb{E}[z_\lambda] = \mathbb{E}[z_\lambda z_{\lambda'}] = 0. 
\]
The function $Z(t)$ then becomes a stochastic process with zero mean and correlations given by Eq. \eqref{eq: noise autocorrelation}. This allows Eq. \eqref{eq: schrodinger equation extended vector} to be interpreted as the stochastic Schrödinger equation \eqref{eq: SSE (I)} in the main text, where the dependence on the random variables was omitted for notational simplicity.

\section{Sampling the noise $Z(t)$}\label{Sampling Z}

On a time grid, the bath-correlation function $\alpha(t,s)$ becomes a two-dimensional matrix $\alpha_{nm} = \alpha(t_n, s_m)$, which we subsequently diagonalize:
\begin{equation}
\alpha_{nm} = \sum_k \lambda_k Y_{n}^{(k)} (Y_{m}^{(k)})^*,
\end{equation}
where $\lambda_k$ are eigenvalues and $Y_{n}^{(k)}$ are the corresponding eigenvectors. Since the BCF is positive semi-definite, $\lambda_k\geq 0$. The eigenvalues $\lambda_k$ and eigenvectors $Y_{n}^{(k)}$ are then used to construct a discretized version of the noise process $Z(t)$:
\begin{equation}
    Z(t_n) = \sum_k \sqrt{\lambda_k} Y_{n}^{(k)} \varepsilon_k,
\end{equation}
where $\varepsilon_k$ are complex Gaussian random variables with zero mean and correlations:
\begin{equation}
    \mathbb{E}[\varepsilon_k \varepsilon_\ell] = 0, 
    \quad\quad
    \mathbb{E}[\varepsilon_k \varepsilon_\ell^*] = \delta_{k\ell}.
\end{equation}
By construction, $Z(t_n)$ has zero mean and correlations $\langle Z(t_n) Z(s_m) \rangle = 0$ and $\langle Z(t_n) Z^*(s_m) \rangle = \alpha_{nm}$.

\subsection*{Special case of $f_j(t)= g_j(t)$}\label{Sampling Z special case}

When $f_j(t) = g_j(t)$, each effective mode can be associated with an independent Ornstein-Uhlenbeck process $z_j(t)$, allowing $Z(t)$ to be written as the following combination:
\begin{equation}\label{eq: noise ansatz}
Z(t) = \sum_{j=1}^{N} f_j(t) z_j(t).
\end{equation}
Each $z_j(t)$ has zero mean and satisfies the correlations
\begin{align}
&\mathbb{E}[z_i(t) z_j(s)] = 0, &
\mathbb{E}[z_i(t) z_j^*(s)] = \delta_{ij} \alpha_j(t-s),
\end{align}
where $\alpha_j(\tau)$ is the exponential function defined in Eq. \eqref{eq: exponential kernel}. 
The Ornstein-Uhlenbeck processes $z_j(t)$ are generated by solving the stochastic differential equations:
\begin{equation}\label{eq: equations for OU processes}
\frac{dz_j(t)}{dt} = -\Gamma_j z_j(t) + \Gamma_j S_j(t),
\end{equation}
with initial conditions $z_j(0) = \xi_j \sqrt{\Gamma_j/2}$, where $\xi_j$ are complex Gaussian random variables with zero mean and correlations $\mathbb{E} [\xi_i \xi_j ] = 0$ and $\mathbb{E} [\xi_i \xi^*_j ] = \delta_{ij}$. The functions $S_i(t)$ represent complex Gaussian white noise with zero mean and correlations:
\begin{align}
&\mathbb{E}[S_i(t) S_j(s)] = 0, & \mathbb{E}[S_i(t) S_j^*(s)] = \delta_{ij} \delta(t - s).
\end{align}

This formulation allows $Z(t)$ to be generated concurrently with the solution of the stochastic Schrödinger equation, without the need to store the full time history of $Z(t)$.

\section{Pseudomode stochastic Schrödinger equation (PSSE)}\label{sec: PSSE}

The density matrix $\rho'(t)$ satisfying Eq. \eqref{eq: pseudomode master equation} can be found as an ensemble average over stochastic pure states
\begin{equation}
\hat{\rho}'(t) = \mathbb{E}\big[|\Psi'(t)\rangle\langle\Psi'(t)|\big].
\end{equation}
The reduced density matrix $\rho_S(t)$ is calculated by tracing out the pseudomode degrees of freedom according to Eq. \eqref{eq: physical density matrix via pseudomode}. The initial condition $|\Psi'(0)\rangle$ is given by:
\begin{equation}
\label{Eq: initial condition SSE}
|\Psi'(0)\rangle = |\psi(0)\rangle \otimes |\mathbf{0}\rangle,
\end{equation}
where the statistics of $|\psi(0)\rangle$ reproduce the initial density matrix $\hat{\rho}_S(0)$ upon averaging.

This stochastic unraveling of Eq. \eqref{eq: pseudomode master equation} leads to the following Itô stochastic differential equation driven by white noise \cite{Disi1998}:
\begin{multline}\label{eq: SSE (II)}
\frac{d|\Psi'(t)\rangle}{dt}  = \Big[ - \sum_{j=1}^{N} \Gamma_j \hat{c}_j^\dag \hat{c}_j - \frac{i}{\hbar} \hat{H}_{\text{eff}}(t) \\
+\sum_{j=1}^{N} \sqrt{2\Gamma_j} S_j^*(t) \hat{c}_j \Big] |\Psi'(t)\rangle,
\end{multline}
where $S_i(t)$ are complex Gaussian white noise terms with zero mean and correlations:
\begin{align}
\mathbb{E}[S_i(t) S_j(s)] = 0, \quad \mathbb{E}[S_i(t) S_j^*(s)] = \delta_{ij} \delta(t-s).
\end{align}
Note that when the rates $\Gamma_j$ are close to zero, the overall state is close to a pure state, which makes the statistical sampling less relevant, thereby leading to better convergence.

Convergence is greatly enhanced by means of the Girsanov transformation, which normalizes contributions from different trajectories \cite{Disi1998}. After the transformation, the stochastic unraveling takes the form:
\begin{equation}
    \hat{\rho}'(t) = \mathbb{E} \left[ \dfrac{|\tilde{\Psi}'(t)\rangle\langle \tilde{\Psi}'(t)|}{\langle\tilde{\Psi}'(t)|\tilde{\Psi}'(t)\rangle} \right],
\end{equation}
where $|\tilde{\Psi}'(t)\rangle$ evolves under a nonlinear stochastic Schrödinger equation:
\begin{multline}\label{eq: SSE (II) nonlinear}
   \dfrac{\partial |\tilde{\Psi}'(t)\rangle}{\partial t} = \Big[ - \sum_{j=1}^{N}  \Gamma_j \hat{c}_j^\dag \hat{c}_j -\dfrac{i}{\hbar} \hat{H}_{\text{eff}}(t) 
     \\
     + \sum_{j=1}^{N} \sqrt{2\Gamma_j} \tilde{S}_j^*(t) \hat{c}_j \Big] |\tilde{\Psi}'(t)\rangle.
\end{multline}
Here, we introduce the shifted noise terms:
\begin{equation}
    \tilde{S}^*_j(t) = S^*_j(t) + \sqrt{2\Gamma_j} \dfrac{\langle \tilde{\Psi}'(t)  | \hat{c}^\dag_i | \tilde{\Psi}'(t)  \rangle}{\langle \tilde{\Psi}'(t) |  \tilde{\Psi}'(t) \rangle}.
\end{equation}
Equation \eqref{eq: SSE (II) nonlinear} is used in numerical examples in Sec. \ref{sec: Strunz single mode}.

\section{Error estimation for master equations}\label{sec: error ME}

Numerical errors in the master equations arise mainly from two sources: (i) the finite time step $h$ used in the integration scheme and (ii) truncation of the hierarchy at levels $n^\text{max}$. We assume these two sources are independent and estimate them separately. Let $\rho_{ij}(t; h, n^\text{max})$ be the reduced density matrix for given $h$ and $n^\text{max}$.

The leading-order time discretization error is estimated using Richardson extrapolation. For this estimation, we set $n^\text{max} = n^{\infty}$, where $n^{\infty}$ is chosen such that further increase changes the result negligibly. For sufficiently small $h$, we assume the error scales polynomially with $h$:
\begin{equation}
    \rho_{ij}(t; h, n^{\infty})
    \approx
    \rho_{ij}(t; n^{\infty})
    +
    C_{ij}(t) h^{p_{ij}(t)},
\end{equation}
where $\rho_{ij}(t; n^{\infty})$ denotes the exact (but unknown) density matrix in the limit $h \to 0$. The absolute value of the last term estimates the time discretization error.

Using solutions for time steps $h$, $2h$, and $4h$, we find the convergence order $p_{ij}(t)$ via
\begin{align}
    p_{ij}(t)
    \approx
    \log_2{\left|
    \dfrac{\rho_{ij}(t; 4h, n^{\infty}) - \rho_{ij}(t; 2h, n^{\infty})}{\rho_{ij}(t; 2h, n^{\infty}) - \rho_{ij}(t; h,n^{\infty})}
    \right|}.
\end{align}
The corresponding global error, assumed to be independent of $n^{\infty}$, is given by
\begin{equation}
\begin{split}
    \Delta_{ij}^{(\text{step})}(t; h) 
    &=
    |C_{ij}(t) h^{p_{ij}(t)}| 
    \\&=
    \dfrac{\big|\rho_{ij}(t; 2h, n^{\infty}) - \rho_{ij}(t; h, n^{\infty})\big|}{2^{p_{ij}(t)} - 1}.
\end{split}
\end{equation}
Our numerical results indicate that $p_{ij}(t)$ is close to $4$, consistent with the expected global accuracy of the RK4 method.

To estimate the truncation error at level $n^\text{max}$, we fix the time step $h$ and subtract the density matrix for $n^\text{max}$ from that for $n^{\infty}$:
\begin{equation}
    \Delta_{ij}^{(\text{trun})}(t; h, n^\text{max})
    \approx
    |\rho_{ij}(t; h, n^\text{max}) - \rho_{ij}(t; h, n^{\infty})|.
\end{equation}
In practice, we observed that $\Delta_{ij}^{(\text{trun})}(t; h, n^\text{max})$ depends only weakly on $h$.

The total numerical error of $\rho_{ij}(t; h, n^\text{max})$ can be found by summing both errors for all matrix elements:
\begin{multline}
    \Delta(t;h,n^\text{max})
    \\
    =
    \sqrt{\sum_{i,j}
    |\Delta_{ij}^{\text{(trun)}}(t; h, n^\text{max})|^2 + \sum_{i,j}|\Delta_{ij}^{\text{(step)}}(t; h)|^2},
\end{multline}
and is plotted in Fig. \ref{fig: ME truncation} (a)-(d). Panels (e)-(h) of the same figure show the root mean square of this quantity:
\begin{equation}
      \sqrt{\dfrac{1}{N_t} \sum_t |\Delta(t; h, n^\text{max})|^2},
\end{equation}
where $N_t$ is the number of stored time points ($1000$ in our simulations).

\section{Error estimation for stochastic methods}\label{sec: error SSE}

Numerical errors in the stochastic methods arise primarily from two sources: (i) a finite number of trajectories and (ii) truncation of the hierarchy at a finite level. The finite time-step error is assumed negligible compared to these two dominant contributions. Let $\rho_{ij}(t; h, n^\text{max})$ denote a single statistical realization of the reduced density matrix with time step $h$ and hierarchy truncation level $n^\text{max}$.

To estimate the error due to hierarchy truncation, we average over $M$ stochastic realizations and compare with a reference density matrix:
\begin{multline}
\label{eq: local trun error stoch}
\Delta(t; h, n^\text{max}, M) \\= \sqrt{\sum_{i,j}\left| \langle \rho_{ij}(t, h, n^\text{max}) \rangle_M - \overline{\rho}_{ij}(t, h, n^{\infty})\right|^2},
\end{multline}
where the reference $\overline{\rho}_{ij}(t, h, n^\text{max})$ is obtained either from a deterministic calculation or from a stochastic calculation averaged over $\overline{M} \gg M$ realizations. The truncation level $n^{\infty}$ is chosen so that further increases do not improve accuracy relative to the sampling-error baseline.

Taking the root mean square of Eq. \eqref{eq: local trun error stoch}, we find:
\begin{equation}
    r(h,n^\text{max},M) = \sqrt{\dfrac{1}{N_t} \sum_t |\Delta(t;h,n^\text{max},M)|^2},
\end{equation}
where $N_t$ is the number of stored time points ($1000$ in our simulations). This metric is shown in Figs. \ref{fig: SSE truncation} (e)-(h) and \ref{fig: SSE truncation Gardiner} (e)-(h). This definition does not explicitly separate statistical sampling error, but under the assumption that fluctuations are independent across time points and dominate other error sources, $r$ can be used to estimate the sampling error. 

To confirm that $r$ also captures sampling errors, we estimated the statistical error (not shown in the figures). When the truncation error is small [constant plateaus in Figs. \ref{fig: SSE truncation} (e)-(h) and \ref{fig: SSE truncation Gardiner} (e)-(h)], the baseline level of $r$ matched the magnitude of statistical fluctuations. Increasing the number of trajectories $M$ would be needed to reduce this baseline.

\bibliography{bib}

\end{document}